\documentclass[reprint,superscriptaddress,showkeys,amsmath,amssymb,aps,pra,longbibliography,floatfix]{revtex4-1}

\usepackage{amsmath}
\usepackage{graphicx}
\usepackage{bm}
\usepackage{dcolumn}
\usepackage{hyperref}

\newcommand{\eps}{\varepsilon}

\newcommand{\kp}{\mathbf{k}\cdot\mathbf{p}}
\newcommand{\Egap}{E_{g}}
\newcommand{\Ep}{E_{P}}
\newcommand{\Esoc}{\Delta_{\text{soc}}}
\newcommand{\Ehf}{E_{\text{HF}}}

\newcommand{\Hkp}{h_{\mathbf{k}\cdot\mathbf{p}}}

\newcommand{\Vconf}{V_{\text{ext}}}

\newcommand{\epseff}{\eps_{\text{eff}}}
\newcommand{\epsopt}{\eps_{\text{opt}}}

\newcommand{\epsin}{\eps_{\text{in}}}
\newcommand{\epsout}{\eps_{\text{out}}}

\newcommand{\Ket}[1]{ | #1 \rangle }

\newcommand{\BraOperKet}[3]{
\langle #1 | #2 | #3 \rangle
}

\newcommand{\Vhf}{V_{\text{HF}}}
\newcommand{\Vdir}{V_{\text{dir}}}
\newcommand{\Vexc}{V_{\text{exc}}}

\begin{document}

\title{Calculation of the biexciton shift in nanocrystals of inorganic perovskites}

\author{T. P. T. Nguyen}
\email{phuctan3108@gmail.com}

\author{S. A. Blundell}
\email{steven.blundell@cea.fr}

\affiliation{Univ.\ Grenoble Alpes, CEA, CNRS, IRIG, SyMMES, F-38000 Grenoble, France}
\author{C. Guet}
\email{cguet@ntu.edu.sg}

\affiliation{Energy Research Institute, Nanyang Technological University, 637141
Singapore}
\affiliation{School of Materials Science and Engineering, Nanyang Technological
University, 639798 Singapore}

\date{\today}

\begin{abstract}
We calculate the shift in emission frequency of the trion and biexciton
(relative to that of the single exciton) for nanocrystals (NCs) of
inorganic perovskites CsPbBr$_{3}$ and CsPbI$_{3}$. The calculations
use an envelope-function $\kp$ model combined with self-consistent
Hartree-Fock and a treatment of the intercarrier correlation energy
in the lowest (second) order of many-body perturbation theory. The
carriers in the trion and biexciton are assumed to have relaxed nonradiatively
to the ground state at the band edge before emission occurs. The theoretical
trion shifts for both CsPbBr$_{3}$ and CsPbI$_{3}$ are found to
be in fair agreement with available experimental data, which include
low-temperature single-dot measurements, though are perhaps systematically
small by a factor of order 1.5, which can plausibly be explained by
a combination of a slightly overestimated dielectric constant and
omitted third- and higher-order terms in the correlation energy. Taking
this level of agreement into account, we estimate that the ground-state
biexciton shift for CsPbBr$_{3}$ is a redshift of order 10--20 meV
for NCs with an edge-length of 12 nm. This value is intermediate among
the numerous high-temperature measurements on NCs of CsPbBr$_{3}$,
which vary from large redshifts of order 100 meV to blueshifts of
several meV. 
\end{abstract}



\maketitle

\section{\label{sec:Introduction} Introduction}

Hybrid organic-inorganic lead halide perovskites such as CH$_{3}$NH$_{3}$PbX$_{3}$
(X = Cl, Br, or I) attracted widespread attention several years ago
on account of their excellent properties for photovoltaic applications
\cite{liu-13-sqd,baikie-13-sqd}. The reported power-conversion efficiencies
have increased rapidly since then and now reach 23.7\% \cite{jiang-17-sqd}.
These high efficiencies are possible in part because the materials
have a high defect tolerance \cite{kang-17-sqd} and very long carrier
diffusion lengths \cite{zhumekenov-16-sqd}.

More recently, nanocrystals (NCs) of all-inorganic lead halide perovskites
CsPbX$_{3}$ (X = Cl, Br, or I) were shown to be outstanding candidates
for light-emitting applications \cite{ProtesescuNanoLett2015}. The
NCs fluoresce strongly, with the emission frequency tunable over the
entire visible range by varying the size of the NCs and their composition
(halide X, including mixtures of different halides) \cite{ProtesescuNanoLett2015}.
The quantum yields obtained are close to 100\% \cite{krieg-18-sqd}.
This has led to important applications of inorganic perovskite NCs
to light-emitting diodes \cite{deng-16-sqd,li-16-sqd}, lasers \cite{pan-15-sqd,yakunin-15-sqd},
and room-temperature single-photon sources \cite{utzat-19-sqd}, among
others.

An important quantity in many light-emitting applications using NCs
is the strength of the exciton-exciton interaction, which causes a
shift in the frequency of light emitted by a biexciton (two confined
excitons) compared to a single exciton. The presence of biexcitons
(or, more generally, of multiexcitons) under device conditions can
reduce the frequency purity of the emitted light, depending on the
size of the shift. The biexciton shift plays a critical role in lasers
based on NCs of II-VI semiconductors such as CdSe, where the small
biexciton redshift is instrumental in creating a population inversion
on the biexciton-to-exciton transition where lasing occurs \cite{nanda-07-sqd}.
It might also be possible to generate polarization-entangled photon
pairs from the biexciton-exciton cascade $\Ket{X\!X}\rightarrow\Ket{X}\rightarrow\Ket{0}$
in NCs of CsPbBr$_{3}$ \cite{utzat-19-sqd}, for which it would help
to understand the energetics of the biexciton decay.

However, the biexciton shift in NCs of CsPbBr$_{3}$ is at present
poorly understood. Many measurements exist \cite{WangAdvMater2015,makarov-16-sqd,CastanedaACSNano2016,AneeshACSjpcc2017,shulenberger-19-sqd,ashner-19-sqd}
that largely contradict one another for reasons that are still controversial,
with reported values of the biexciton shift varying from large redshifts
\cite{CastanedaACSNano2016} of order 100~meV to a recently reported
small blueshift \cite{ashner-19-sqd} of order a few meV.

To help understand this issue, we present here calculations of the
biexciton shift in NCs of CsPbI$_{3}$ and CsPbBr$_{3}$ using a multiband
$\kp$ envelope-function approach, combined with many-body perturbation
theory (MBPT). We assume that the biexciton has relaxed nonradiatively
(by rapid phonon emission) to its ground state at the band edge before
emitting, which enables us to construct a detailed microscopic theory
of the multicarrier correlations responsible for the shift. Our results
suggest that the biexciton shift under these conditions is a redshift
having a value that is intermediate among the available measurements
on NCs of CsPbBr$_{3}$.

The plan of the paper is the following. In Sec.~\ref{sec:Methods}
we outline our formalism. We treat the confined carriers as an `artificial
atom' using methods of MBPT from atomic physics and quantum chemistry
\cite{LindgrenMorrison,ShavittBartlett}. The first step is a self-consistent
Hartree-Fock (HF) model of the confined carriers; then we apply the
leading correlation correction from second-order MBPT. Our basic envelope-function
model is discussed in Sec.~\ref{sec:model}, the HF method in Sec.~\ref{sec:HFApproximation},
and the correlation energy in Sec.~\ref{sec:MBPTFormalism}. For
reasons of computational efficiency, we use a spherical basis set
in the MBPT calculations. This leads to extensive formulas for the
various terms involving radial integrals and angular factors, which
can be derived using standard methods of angular-momentum theory \cite{LindgrenMorrison,Brink&Satchler,Edmonds}.
These detailed formulas will be presented elsewhere.

These methods are then applied to NCs of inorganic perovskites in
Sec.~\ref{sec:ResultsDiscussions}. A difficulty with these materials,
which have only recently become the subject of intensive research,
is that many of their properties are at present poorly understood.
Even some basic properties, such as the effective masses of the valence
and conduction bands, are uncertain. We discuss the available data
and the parameters that we assume in our model in Sec.~\ref{sec:parameters}.
Next, in Sec.~\ref{sec:ClosedShell}, we apply our approach to the
trion and biexciton shift in NCs of CsPbI$_{3}$ and CsPbBr$_{3}$.
Both these shifts are dominated by intercarrier correlation effects,
the mean-field (HF) contribution largely canceling \cite{ZungerCorrPRB2001}.
As we shall see, the calculations of the correlation energy for trions
and biexcitons are very closely related, so that the data on trion
shifts provide a very useful additional check on our calculation of
the biexciton shift. Our conclusions are given in Sec.~\ref{sec:Conclusions}.

\section{\label{sec:Methods} Formalism}

\subsection{\label{sec:model}Model}

Our approach is based on an envelope-function formalism \cite{Kira&Koch}
for a system of carriers (holes and electrons) confined in a potential
$\Vconf$, with the bulk band structure described by a $\kp$ Hamiltonian
$\Hkp$ and screened Coulomb interactions among the carriers. The
total Hamiltonian in the space of envelope functions is 
\begin{eqnarray}
H & = & \sum_{ij}\{i^{\dagger}j\}\BraOperKet{i}{\Hkp+\Vconf}{j}\nonumber \\
 &  & {}+\frac{1}{2}\sum_{ijkl}\{i^{\dagger}j^{\dagger}lk\}\BraOperKet{ij}{g_{12}}{kl}\,,\label{eq:hamiltonian}
\end{eqnarray}
where $\{i_{1}^{\dagger}i_{2}^{\dagger}\ldots j_{1}j_{2}\ldots\}$
is a normally ordered product of creation (and absorption) operators
for electron envelope states $i_{1},i_{2},\ldots$ (and $j_{1},j_{2},\ldots$),
which span the conduction bands (CBs) and valence bands (VBs) included
in the calculation. The Coulomb interaction $g_{12}$ in envelope-function
approaches is given generally by a sum of long-range (LR) and short-range
(SR) terms \cite{Knox,PikusJETP1971}. Here we will consider only
the LR part (we use atomic units throughout) 
\begin{eqnarray}
g_{12} & = & \frac{1}{\epsin|\mathbf{r}_{1}-\mathbf{r}_{2}|}\,,\label{eq:lrcoul}
\end{eqnarray}
where $\epsin$ is the dielectric constant of the NC material appropriate
to the length scale $L_{\text{dot}}$ of the nanostructure (see Sec.~\ref{sec:parameters}).
The LR Coulomb interaction is in principle modified by the mismatch
with the dielectric constant $\epsout$ of the surrounding medium,
which leads to induced polarization charges at the interface, although
we will not consider this effect in the present paper.

Even though perovskite NCs are generally cuboid, we use a basis of
envelope states $i,j,\ldots$, etc., in Eq.~(\ref{eq:hamiltonian})
appropriate to spherical symmetry. This is done for reasons of computational
efficiency. In a spherical basis, the angular integrals can be carried
out analytically and the remaining radial integrals are one-dimensional.
It is also possible to sum over the magnetic substates of the basis
states analytically \cite{LindgrenMorrison,Brink&Satchler}, which
effectively reduces (very substantially) the size of the basis required
in correlation calculations. Although we will not do so in this paper,
nonspherical terms in the Hamiltonian (for example, arising from the
crystal lattice or from the overall shape of the NC) can in principle
be included in later stages of the formalism as perturbations.

To generate a spherical basis, we take the confining potential to
be spherically symmetric. We choose a spherical well with infinite
walls, 
\begin{equation}
\Vconf(r)=\left\{ \begin{matrix}0\text{, if }r<R\\
\infty\text{, otherwise}
\end{matrix}\right.\,.\label{eq:sphericalWell}
\end{equation}
If the NC is a cube with edge-length $L$, the radius $R$ can be
conveniently chosen to satisfy 
\begin{equation}
R=L/\sqrt{3}\,.\label{eq:radiusL}
\end{equation}
To motivate this choice of $R$, we note that at effective-mass level
the eigenvalues of noninteracting electrons in a cubic box are given
by 
\begin{equation}
\epsilon_{\lambda}^{\text{cube}}(n_{x},n_{y},n_{z})=\frac{\pi^{2}}{2m_{\lambda}^{*}L^{2}}(n_{x}^{2}+n_{y}^{2}+n_{z}^{2})\,,\label{eq:eigsCube}
\end{equation}
where $(n_{x},n_{y},n_{z})$ are integers and $m_{\lambda}^{*}$ is
the band effective mass. Thus, the condition~(\ref{eq:radiusL})
ensures that the entire spectrum of `$S$-like' states in a cube ($n_{x}=n_{y}=n_{z}=n$)
coincides exactly with the spectrum of $nS$ states in a sphere, 
\begin{equation}
\epsilon_{\lambda}^{\text{sph}}(n)=\frac{\pi^{2}n^{2}}{2m_{\lambda}^{*}R^{2}}\,.\label{eq:eigsSphere}
\end{equation}
One can also show that the lowest `$P$-like' state in a cube ($n_{x}=2$,
$n_{y}=n_{z}=1$, together with the two other permutations, $n_{x}\leftrightarrow n_{y}$
and $n_{x}\leftrightarrow n_{z}$ \cite{shaw-74-sqd}) has an energy
within 2.3\% of that of the $1P$ state in the equivalent sphere~(\ref{eq:radiusL}),
and that higher-lying `$P$-like' states also have energies within
several percent of their analog in the sphere.

Even though the single-particle energies are in close agreement, wave
functions and therefore matrix elements can still differ between cubic
and spherical confinement. However, in Sec.~\ref{sec:HFApproximation}
we show that the first-order Coulomb energy of the ground-state exciton
differs by only about 1.5\% in the two cases, and the HF energy by
about 0.04\%. In Sec.~\ref{sec:MBPTFormalism}, we estimate that
the error in the correlation energy from using a spherical basis is
about 5\%. Therefore, for the purposes of this paper, the nonspherical
correction term arising from the NC shape is expected to be unimportant.

We consider two $\kp$ models. The first is a $4\times4$ model, which
includes the $s$-like VB and $p_{1/2}$-like CB around the $R$ point
of the Brillouin zone in inorganic perovskite compounds \cite{even-14b-sqd,BeckerNatLett2018}.
The other is an $8\times8$ model including additionally the $p_{3/2}$-like
CB, which lies about 1~eV above the $p_{1/2}$-like CB at the $R$
point \cite{even-14b-sqd,YuSciRep2016,BeckerNatLett2018}. Including
the $p_{3/2}$-like CB in this way leads to a small correction to
correlation energies at the 1\% level (see Sec.~\ref{sec:MBPTFormalism}).

For spherical confinement, the angular part of an envelope function
with orbital angular momentum $l$ couples to a Bloch function with
Bloch angular momentum $J$ (here $J=1/2$ or 3/2) to give a state
with total angular momentum $(F,m_{F})$ \cite{ekimov-93-sqd}, which
we denote by a basis vector $\Ket{(l,J)Fm_{F}}$. In the $8\times8$
model, the total wave function (including envelope and Bloch functions)
can then be written as a sum of four components \cite{ekimov-93-sqd},
\begin{align}
 & \Ket{\eta Fm_{F}}=\nonumber \\
 & \quad\frac{g_{s}(r)}{r}\Ket{(l+1,1/2)Fm_{F}}+\frac{\bar{g}_{p}(r)}{r}\Ket{(\bar{l},1/2)Fm_{F}}\nonumber \\
 & \quad{}+\frac{g_{p}(r)}{r}\Ket{(l,3/2)Fm_{F}}+\frac{f_{p}(r)}{r}\Ket{(l+2,3/2)Fm_{F}}\,.\label{eq:4compState}
\end{align}
Here $g_{s}(r)$ and $\bar{g}_{p}(r)$ are the radial envelope functions
for the $s$-like and $p_{1/2}$-like bands, respectively, while $g_{p}(r)$
and $f_{p}(r)$ apply to the $p_{3/2}$-like band. These last two
terms are absent in the $4\times4$ model. The allowed values of the
angular momenta $l$ and $\bar{l}$ follow from angular-momentum and
parity selection rules \cite{ekimov-93-sqd}. We solve for the radial
functions and eigenvalues of the single-particle states in the presence
of a Hartree-Fock potential (Sec.~\ref{sec:HFApproximation}) using
a generalization of the method of Ref.~\cite{ekimov-93-sqd}.

For states in the $s$-like VB, the term involving $g_{s}(r)$ in
Eq.~(\ref{eq:4compState}) is typically the large component of the
wave function, while the other terms are small components representing
the admixture of CB states into the VB states due to the finite range
of the confining potential $\Vconf$ and the $\kp$ interaction. In
the CB states, the role of the small and large components are interchanged.
The presence of the small components allows the formalism to pick
up the leading $\kp$ corrections arising from the coupling of the
VB and CB.

\subsection{\label{sec:HFApproximation} Hartree-Fock}

The first step in the correlation calculation for a general excitonic
system with $N_{e}$ electrons and $N_{h}$ holes is to solve the
self-consistent HF equations including exact exchange \cite{LindgrenMorrison,ShavittBartlett}.
The HF potential will then be used to define the single-particle states
of the many-body procedure discussed in Sec.~\ref{sec:MBPTFormalism}.

For an occupied state $\Ket{a}$ (either a hole or an electron), the
HF equation is 
\begin{equation}
\left(\Hkp+\Vconf+\Vhf^{\text{av}}\right)\Ket{a}=\epsilon_{a}\Ket{a}\,,\label{eq:HFeqn}
\end{equation}
where the HF potential $\Vhf^{\text{av}}$ is given by a sum of direct
and exchange terms, $\Vhf^{\text{av}}=\Vdir+\Vexc$, with 
\begin{eqnarray}
\BraOperKet{i}{\Vdir}{a} & = & \sum_{b}^{\text{occ}}e_{b}q_{b}^{a}\BraOperKet{ib}{g_{12}}{ab}\,,\label{eq:HFdir}\\
\BraOperKet{i}{\Vexc}{a} & = & -\sum_{b}^{\text{occ}}e_{b}q_{b}^{a}\BraOperKet{ib}{g_{12}}{ba}\,,\label{eq:HFexc}
\end{eqnarray}
where the sum is over all occupied (or partially occupied) states.
Here $e_{b}$ is a charge-related parameter, with $e_{b}=1$ for electrons
and $e_{b}=-1$ for holes. (We are using the convention that eigenvalues
$\epsilon_{a}$ refer to electron states, even though the states may
be `occupied' by a hole with an energy $-\epsilon_{a}$.)

The usual HF potential with $q_{b}^{a}=1$ in Eqs.~(\ref{eq:HFdir})
and (\ref{eq:HFexc}) is generally only a scalar operator for closed-shell
systems. Since we wish to create a spherical basis for open-shell
systems as well, we employ instead a \emph{configuration-averaged
HF} \cite{LindgrenMorrison}, in which the configuration-averaging
weights $q_{b}^{a}$ are given by 
\begin{equation}
q_{b}^{a}=\left\{ \begin{array}{cc}
n_{B}/g_{B} & \quad b\notin A\\
(n_{B}-1)/(g_{B}-1) & \quad b\in A
\end{array}\right.\,.\label{eq:configWeights}
\end{equation}
Here $A$ or $B$ denotes the shell containing the states $a$ or
$b$, respectively, $n_{B}$ is the occupation number of shell $B$,
and $g_{B}$ is the degeneracy (maximum occupation) of shell $B$.
For a closed-shell system, $n_{B}=g_{B}$ for all shells and then
all weights $q_{b}^{a}=1$. The configuration-averaged HF equations~(\ref{eq:HFeqn})
for a spherically symmetric $\Vconf$ can now be reduced to a set
of radial HF equations following standard procedures \cite{LindgrenMorrison}.

The configuration-averaged HF energy of the excitonic system is 
\begin{eqnarray}
E_{\text{HF}}^{\text{av}} & = & \sum_{a}^{\text{occ}}e_{a}q_{a}\BraOperKet{a}{\Hkp+\Vconf}{a}\nonumber \\
 &  & {}+\frac{1}{2}\sum_{a}^{\text{occ}}e_{a}q_{a}\BraOperKet{a}{\Vhf^{\text{av}}}{a}\,,\label{eq:Ehfav}
\end{eqnarray}
where 
\begin{equation}
q_{a}=n_{A}/g_{A}\label{eq:qaWeight}
\end{equation}
is the fractional occupation of shell $A$ (where $a\in A$). Conventionally
we define the zero of the band-structure energy to be the VB maximum.
Then we can decompose $E_{\text{HF}}^{\text{av}}$ into different
physical contributions as 
\begin{equation}
E_{\text{HF}}^{\text{av}}=E_{\text{band}}+E_{\text{conf}}+E_{\text{Coul}}\,,\label{eq:decompEhf}
\end{equation}
where $E_{\text{band}}=N_{e}E_{g}$ is the `band energy' ($\Egap$
is the gap between the $s$-like VB and the $p_{1/2}$-like CB) and
$E_{\text{conf}}$ is the confinement energy, 
\begin{equation}
E_{\text{conf}}=\sum_{a}^{\text{occ}}e_{a}q_{a}\BraOperKet{a}{\Hkp}{a}-E_{\text{band}}\,.\label{eq:Econf}
\end{equation}
One can also define an energy of interaction with the external potential,
$E_{\text{ext}}=\sum_{a}^{\text{occ}}e_{a}q_{a}\BraOperKet{a}{\Vconf}{a}$,
although here $E_{\text{ext}}\equiv0$ because of our simple choice
of potential~(\ref{eq:sphericalWell}). The Coulomb energy is 
\begin{equation}
E_{\text{Coul}}=\frac{1}{2}\sum_{a}^{\text{occ}}e_{a}q_{a}\BraOperKet{a}{\Vhf^{\text{av}}}{a}\,,\label{eq:Ecoul}
\end{equation}
which can be further decomposed into direct and exchange terms using
Eqs.~(\ref{eq:HFdir}) and~(\ref{eq:HFexc}). Example calculations
showing these energy contributions for a NC of CsPbBr$_{3}$ are given
in Table~\ref{tab:HF}. Note that the exchange energy for a single
exciton is very small; this contribution can be shown to be formally
of order $(L_{\text{atom}}/L_{\text{dot}})^{2}$, where $L_{\text{atom}}$
is the interatomic length scale.

\begin{table}
\caption{\label{tab:HF}Hartree-Fock calculation for a ground-state single
exciton ($X$), negative trion ($X^{-}$), and biexciton ($X\!X$)
confined in a NC of CsPbBr$_{3}$ with edge-length $L=9$~nm, using
the material parameters in Table~\protect\ref{tab:parameters} (and
$\Ep=20$~eV). $E_{\text{band}}$ is the band energy, $E_{\text{conf}}$
the confinement energy, $E_{\text{dir}}$ and $E_{\text{exc}}$ are
the direct and exchange Coulomb energy, respectively, $E_{\text{Coul}}$
is the total Coulomb energy, $E_{\text{Coul}}=E_{\text{dir}}+E_{\text{exc}}$,
and $\Ehf=E_{\text{band}}+E_{\text{conf}}+E_{\text{Coul}}$ is the
total HF energy.}

\begin{ruledtabular}
\begin{tabular}{lddd}
 & \multicolumn{1}{c}{$X$ (eV)} & \multicolumn{1}{c}{$X^{-}$ (eV)} & \multicolumn{1}{c}{$X\!X$ (eV)}\\
\hline 
$E_{\text{band}}$  & 2.3420  & 4.6840  & 4.6840 \\
$E_{\text{conf}}$  & 0.1036  & 0.1556  & 0.2071 \\
$E_{\text{dir}}$  & -0.0699  & -0.0069  & 0.0000 \\
$E_{\text{exc}}$  & 0.0003  & -0.0655  & -0.1385 \\
$E_{\text{Coul}}$  & -0.0696  & -0.0724  & -0.1385 \\
$\Ehf$  & 2.3760  & 4.7671  & 4.7526 \\
\end{tabular}
\end{ruledtabular}

\end{table}

To study the dependence of the HF energy on the shape of the NC (sphere
or cube), consider the $1S_{e}$-$1S_{h}$ ground state of a single
exciton. In the effective-mass limit, the noninteracting $1S$ states
(electron or hole) have wave functions 
\begin{equation}
\psi_{1S}^{\text{cube}}(\mathbf{r})=\sqrt{\frac{8}{L^{3}}}\cos\left(\frac{\pi x}{L}\right)\cos\left(\frac{\pi y}{L}\right)\cos\left(\frac{\pi z}{L}\right)\label{eq:psiCube}
\end{equation}
for cubic confinement, and 
\begin{equation}
\psi_{1S}^{\text{sph}}(r)=\frac{1}{\sqrt{2\pi R}}\frac{1}{r}\sin\left(\frac{\pi r}{R}\right)\label{eq:psiSphere}
\end{equation}
for spherical confinement. The confinement (kinetic) energy of the
the $1S_{e}$-$1S_{h}$ exciton at this level of approximation follows
from Eqs.~(\ref{eq:radiusL})--(\ref{eq:eigsSphere}) to be 
\begin{equation}
E_{\text{conf}}^{(1)}=\frac{3\pi^{2}}{2L^{2}}\left(\frac{1}{m_{e}^{*}}+\frac{1}{m_{h}^{*}}\right)\label{eq:Ekin1S}
\end{equation}
for both the cube and the equivalent sphere~(\ref{eq:radiusL}).
The first-order Coulomb energy can be obtained by inserting the wave
functions (\ref{eq:psiCube}) and (\ref{eq:psiSphere}) into Eqs.~(\ref{eq:HFdir})
and (\ref{eq:Ecoul}), and neglecting the exchange term. This gives
\begin{equation}
E_{\text{Coul}}^{(1)}=-\frac{\xi}{\epsin L}\,,\label{eq:E1coul}
\end{equation}
where, after numerical integration, we find $\xi\approx4.389$~eV\,nm
(for a cube) and $\xi\approx4.455$~eV\,nm (for a sphere). Thus,
the Coulomb energy differs by about 1.5\% between the cube and the
equivalent sphere. From Eqs.~(\ref{eq:decompEhf}), (\ref{eq:Ekin1S}),
and (\ref{eq:E1coul}) we then find that, for the parameters used
in Table~\ref{tab:HF}, the HF energy of the single exciton at this
level of approximation is $\Ehf=2.3858$~eV (for a cube) and $\Ehf=2.3848$~eV
(for a sphere), a difference of only 0.04\%. Finally, one sees from
Table~\ref{tab:HF} that the HF energy for a sphere changes by 0.4\%
from this value upon incorporating $\kp$ corrections (with $\Ep=20$~eV)
and iterating the HF equations to self-consistency.

\subsection{\label{sec:MBPTFormalism} Correlation energy}

From the point of view of MBPT \cite{LindgrenMorrison,ShavittBartlett},
the HF energy of a closed-shell system is correct through first order,
$\Ehf=E^{(0)}+E^{(1)}$, where $E^{(0)}=\sum_{a}^{\text{occ}}e_{a}\epsilon_{a}$
is the sum of the single-particle eigenvalues of the occupied HF states,
and $E^{(1)}$ is the first-order correction of the residual Coulomb
interaction. The configuration-averaged HF energy~(\ref{eq:Ehfav})
of an open-shell system is similar, but gives the energy of the center
of gravity of the configuration multiplet, again correct through first
order in MBPT \cite{LindgrenMorrison}. The higher-order corrections
to the energy, $E_{\text{corr}}=E^{(2)}+E^{(3)}+\ldots$, are referred
to as the \emph{correlation energy}.

In this paper we will consider only the second-order energy, $E_{\text{corr}}\approx E^{(2)}$.
For atoms and molecules, $E^{(2)}$ typically accounts for about 75\%
or more of the total correlation energy (depending on the system)\cite{LindgrenMorrison,ShavittBartlett}
and usually $E^{(2)}/E_{\text{corr}}<1.$ This approximation has the
merit of simplicity. Using the spherical basis~(\ref{eq:4compState}),
$E^{(2)}$ for the excitonic systems considered here can be converged
to an accuracy of a fraction of a percent in about 1~s or less on
a single processing core.

\begin{figure}[tb]
\includegraphics[scale=0.035]{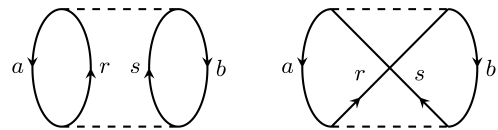} \caption{Closed-shell second-order correlation energy: direct (on the left)
and exchange (on the right).}
\label{ClosedShell2ndorder} 
\end{figure}

The second-order energy for a closed-shell atom or molecule in a HF
potential is given \cite{LindgrenMorrison,ShavittBartlett} by the
many-body diagrams in Fig.~\ref{ClosedShell2ndorder}. To apply this
approach to an excitonic system of holes and electrons, we will effectively
consider the electrons and holes to be different species of particle
and evaluate the diagram for the mixed system \footnote{By so doing, we exclude MBPT terms involving the creation of electron-hole
pairs, or \emph{virtual excitons}, in the intermediate states. Such
terms are however naturally suppressed by the small Coulomb matrix
elements associated with them.}. Thus, the lines directed downward in Fig.~\ref{ClosedShell2ndorder}
($a$ and $b$) correspond to \emph{occupied} states (either holes
or electrons) while upward-directed lines ($r$ and $s$) correspond
to \emph{unoccupied} states (either holes or electrons). The total
second-order energy is given by 
\begin{equation}
E^{(2)}=\frac{1}{2}\sum_{abrs}D_{abrs}^{(2)}\,,\label{eq:E2general}
\end{equation}
where 
\begin{equation}
D_{abrs}^{(2)}=\frac{\BraOperKet{ab}{g_{12}}{rs}(\BraOperKet{rs}{g_{12}}{ab}-\BraOperKet{rs}{g_{12}}{ba})}{\omega_{a}+\omega_{b}-\omega_{r}-\omega_{s}}\,,\label{eq:D2general}
\end{equation}
and $\omega_{i}=\epsilon_{i}$ for electrons and $\omega_{i}=-\epsilon_{i}$
for holes, since the single-particle energies must now apply to each
particle type. Decomposing $E^{(2)}$ explicitly into electron and
hole contributions gives 
\begin{equation}
E^{(2)}=E_{ee}^{(2)}+E_{hh}^{(2)}+E_{eh}^{(2)}\,,\label{eq:E2decomposed}
\end{equation}
where 
\begin{align*}
E_{ee}^{(2)} & =\frac{1}{2}\sum_{abrs}^{\text{elec}}D_{abrs}^{(2)}\,, & E_{hh}^{(2)} & =\frac{1}{2}\sum_{abrs}^{\text{hole}}D_{abrs}^{(2)}\,,
\end{align*}
\begin{align}
E_{eh}^{(2)} & =\sum_{\begin{subarray}{c}
ar\,\text{(elec)}\\
bs\,\text{(hole)}
\end{subarray}}D_{abrs}^{(2)}\,.\label{eq:E2terms}
\end{align}
The terms $E_{ee}^{(2)}$ and $E_{hh}^{(2)}$ correspond to the correlation
energy of the separate electron and hole subsystems, respectively.
The third term $E_{eh}^{(2)}$ is a cross-term, involving single excitations
of both the electron and hole subsystems.

A general excitonic system with $N_{e}$ electrons and $N_{h}$ holes
may contain open shells. An approximate formula for the correlation
energy in this case may be found by inserting configuration-averaging
weights for the occupied or partially occupied shells of $a$ and
$b$ into the closed-shell formula~(\ref{eq:E2general}), following
the same argument used for the configuration-averaged HF energy \cite{LindgrenMorrison}.
Equation~(\ref{eq:E2general}) is then modified by 
\begin{equation}
D_{abrs}^{(2)}\rightarrow q_{a}q_{b}^{a}D_{abrs}^{(2)}\,,\label{eq:D2av}
\end{equation}
where $q_{b}^{a}$ and $q_{a}$ are given by Eqs.~(\ref{eq:configWeights})
and~(\ref{eq:qaWeight}), respectively.

To evaluate the sums over states in Eq.~(\ref{eq:E2general}), we
create a basis set of single-particle states in the HF potential~(\ref{eq:HFeqn})
up to a high energy cutoff. This basis set contains the occupied or
partially occupied states $a$ and $b$, which contribute to the HF
potential, together with unoccupied (excited) states $r$ and $s$.
For the calculations on single excitons, trions, and biexcitons presented
in this paper, we take $q_{b}^{a}$ for any \emph{unoccupied} state
$a$ to be $q_{b}^{c}$, where $c$ is chosen to be the $1S_{e}$
state (for electrons) or the $1S_{h}$ state (for holes). This choice
forces all $S$-wave excited states in the basis set to be orthogonal
to the occupied $1S_{e}$ and $1S_{h}$ states (as required).

\begin{table}
\caption{\label{tab:correlation}Second-order correlation energy for a ground-state
single exciton ($X$), negative trion ($X^{-}$), and biexciton ($X\!X$)
confined in a NC of CsPbBr$_{3}$ with edge-length $L=9$~nm, using
the material parameters in Table~\protect\ref{tab:parameters} (and
$\Ep=20$~eV). $E_{ee}^{(2)}$, $E_{hh}^{(2)}$, and $E_{eh}^{(2)}$
are the electron term, the hole term, and the electron-hole cross-term,
respectively, given by Eq.~(\protect\ref{eq:E2terms}). Direct (dir)
and exchange (exc) terms are shown separately; the exchange term from
$E_{eh}^{(2)}$ is negligible. First three columns: $4\times4$ $\kp$
model; last column: $8\times8$ $\kp$ model. Units: meV.}

\begin{ruledtabular}
\begin{tabular}{ldddd}
 & \multicolumn{3}{c}{$4\times4$ $\kp$} & \multicolumn{1}{c}{$8\times8$ $\kp$}\\
 & \multicolumn{1}{c}{$X$} & \multicolumn{1}{c}{$X^{-}$} & \multicolumn{1}{c}{$X\!X$} & \multicolumn{1}{c}{$X$ }\\
\hline 
$E_{ee}^{(2)}$ (dir)  & 0.00  & -8.44  & -8.41  & 0.00 \\
$E_{ee}^{(2)}$ (exc)  & 0.00  & 4.21  & 4.20  & 0.00 \\
$E_{hh}^{(2)}$ (dir)  & 0.00  & 0.00  & -8.41  & 0.00 \\
$E_{hh}^{(2)}$ (exc)  & 0.00  & 0.00  & 4.20  & 0.00 \\
$E_{eh}^{(2)}$ (dir)  & -6.83  & -10.22  & -16.82  & -7.13 \\
$E^{(2)}$ (total)  & -6.83  & -14.44  & -25.24  & -7.13 \\
\end{tabular}
\end{ruledtabular}

\end{table}

Example calculations of the second-order correlation energy are given
in Table~\ref{tab:correlation}. Note that only the cross-term $E_{eh}^{(2)}$
contributes for a single exciton, since the configuration-averaging
weights in Eq.~(\ref{eq:D2av}) vanish for the other terms. Also,
the electron $E_{ee}^{(2)}$ and hole $E_{hh}^{(2)}$ terms here contribute
equally for the biexciton, because we assume VB-CB symmetry in the
material parameters; this is not true in general. The sums over the
intermediate states $r$ and $s$ are quite rapidly convergent: about
10\% of $E^{(2)}$ arises from the $S$-wave channel, 70\% from the
$P$-wave channel, and 18\% from the $D$- and $F$-wave channels.
In addition, the first three principal quantum numbers of each angular
channel are sufficient to obtain about 98\% of $E^{(2)}$. We note
that the contributions to $E^{(2)}$ presented contain small $\kp$
corrections of about 2\%.

From Table~\ref{tab:correlation}, we see that using the $8\times8$
$\kp$ model modifies the single-exciton correlation energy by only
about 4\% compared to the $4\times4$ model. Actually, most of this
shift is due to the modification of the $\kp$ corrections by the
presence of the $p_{3/2}$-like band. If the calculations are repeated
in the effective-mass limit ($\Ep\rightarrow0$), one finds a difference
of only about 0.1\% between the $8\times8$ and $4\times4$ models,
showing that the excitations into the $p_{3/2}$-like band are not
very significant in themselves (owing to their relatively high excitation
energy). This justifies the use of the $4\times4$ $\kp$ model for
perovskite NCs for the calculation of the correlation energy.

Noting that the dominant intermediate channel is $P$-wave, we estimate
the error in $E^{(2)}$ from using a spherical (not cubic) basis to
be about 5\%, which is the error in the energy denominator associated
with the $1S\rightarrow nP$ excitations for $n=1$--3 (see Sec.~\ref{sec:model}).

For an alternative approach to correlation in a confined excitonic
system with spherical symmetry, see Ref.~\cite{chang-98-sqd}.

\section{\label{sec:ResultsDiscussions}Application to perovskite nanocrystals}

\subsection{\label{sec:parameters}Parameters}

\begin{table}[tb]
\caption{\label{tab:parameters}Parameters used in the calculations. $\Ep^{(1)}$
is the Kane parameter estimated from the $4\times4$ $\mathbf{k}\cdot\mathbf{p}$
model, $\Ep^{(2)}$ from the $8\times8$ $\mathbf{k}\cdot\mathbf{p}$
model. For further explanation, see Sec.~\protect\ref{sec:parameters}.}

\begin{ruledtabular}
\begin{tabular}{ldd}
 & \multicolumn{1}{c}{CsPbBr$_{3}$} & \multicolumn{1}{c}{CsPbI$_{3}$}\\
\hline 
$\Egap$ (eV)  & 2.342\footnotemark[1]  & 1.723\footnotemark[1] \\
$\mu^{*}$ ($m_{0}$)  & 0.126\footnotemark[1]  & 0.114\footnotemark[1] \\
$m_{e}^{*},m_{h}^{*}$ ($m_{0}$)  & 0.252  & 0.228\\
$\Esoc$ (eV)  & 1.0\footnotemark[2]  & 1.0\footnotemark[2]\\
$\epseff$  & 7.3\footnotemark[1]  & 10.0\footnotemark[1] \\
$\epsopt$  & 5.3\footnotemark[3]  & 4.8\footnotemark[4]\\
$\Ep^{(1)}$ (eV)  & 27.9  & 22.7\\
$\Ep^{(2)}$ (eV)  & 16.4  & 13.9 \\
\end{tabular}

\footnotetext[1]{Ref.~\cite{YangACSEnergyLett2017}}

\footnotetext[2]{Ref.~\cite{YuSciRep2016}}

\footnotetext[3]{Ref.~\cite{DirinEpsVersusLambdaACSchemmater2016},
at a wavelength of 600~nm.}

\footnotetext[4]{Ref.~\cite{SinghEpsVersusLambdaJjtice2019}, at
a wavelength of 600~nm.} 
\end{ruledtabular}

\end{table}

The material parameters that we use for CsPbBr$_{3}$ and CsPbI$_{3}$
are summarized in Table~\ref{tab:parameters}. The bulk parameters
$\mu^{*}$ and $\epseff$ are taken from Ref.~\cite{YangACSEnergyLett2017}
and apply to the orthorhombic phase of CsPbBr$_{3}$ and the cubic
phase of CsPbI$_{3}$ at cryogenic temperatures \cite{CottinghamChemComm2016,StoumposACScg2013,HirotsuJPSJ1974}.
Although the reduced mass $\mu^{*}=m_{e}^{*}m_{h}^{*}/(m_{e}^{*}+m_{h}^{*})$
has been measured \cite{YangACSEnergyLett2017} by magneto-transmission
techniques, the individual effective masses of electron $m_{e}^{*}$
and hole $m_{h}^{*}$ are unknown. Evidence from experiment \cite{SumTCNatCom2017}
and first-principles calculations \cite{BeckerNatLett2018,ProtesescuNanoLett2015,UmariSciRep2014}
suggests, however, that $m_{e}^{*}$ and $m_{h}^{*}$ are approximately
equal. Here we will assume $m_{e}^{*}=m_{h}^{*}$ exactly ($m_{e}^{*}$
applies to the $p_{1/2}$-like CB, and $m_{h}^{*}$ to the $s$-like
VB, around the $R$ point of the Brillouin zone). The spin-orbit splitting
$\Esoc$ between the $p_{1/2}$-like and the higher-lying $p_{3/2}$-like
band has been measured in Ref.~\cite{YuSciRep2016}.

The `effective' dielectric constant $\epseff$ in Table~\ref{tab:parameters}
is derived \cite{YangACSEnergyLett2017} from the measured binding
energy of the bulk exciton. We also give for comparison values of
the optical dielectric constant $\epsopt$ at a wavelength of 600~nm,
which are somewhat smaller than $\epseff$. The constant $\epseff$
applies to a length scale of order the bulk Bohr radius $a_{B}$,
which is quite close to the size of the NCs that we consider ($2a_{B}=6.1$~nm
for CsPbBr$_{3}$ and $2a_{B}=9.3$~nm for CsPbI$_{3}$, using the
parameters in Table~\ref{tab:parameters}). Therefore we shall use
$\epsin=\epseff$ to screen the LR Coulomb interaction~(\ref{eq:lrcoul})
in the main parts of our calculations of the correlation and exchange
energy.

\begin{figure}
\includegraphics[scale=0.85]{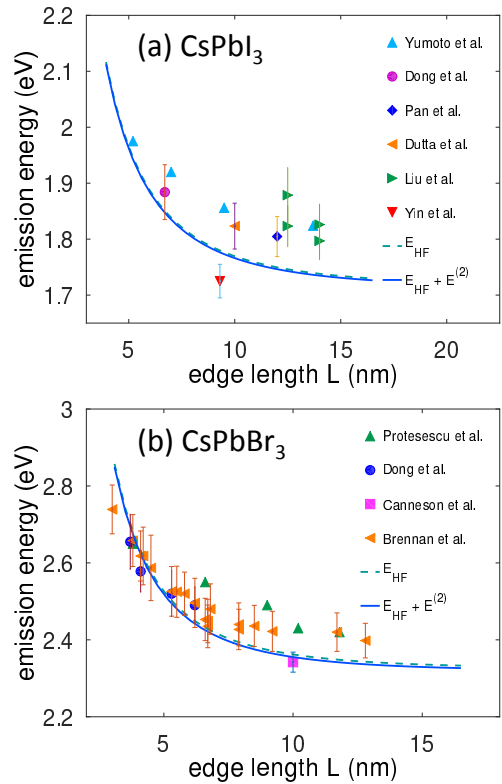}

\caption{\label{fig:PLpeaks}Measured photoluminescence peak energies of NCs
of (a) CsPbI$_{3}$ and (b) CsPbBr$_{3}$ (triangles/squares/circles),
and theoretical single-exciton energy using HF (dashed curve) and
HF plus second-order correlation energy (full curve). Yumoto \emph{et
al}., Ref.~\protect\cite{yumoto-18-sqd}; Dong \emph{et al}., Ref.~\protect\cite{DongNanoLett2018};
Pan \emph{et al}., Ref.~\protect\cite{PanJACS2018}; Dutta \emph{et
al}., Ref.~\protect\cite{Dutta18Angew2018}; Liu \emph{et al}.,
Ref.~\protect\cite{LiuACSNano2017}; Yin \emph{et al}., Ref.~\protect\cite{YinPRL2017};
Protesescu \emph{et al}., Ref.~\protect\cite{ProtesescuNanoLett2015};
Canneson \emph{et al}., Ref.~\protect\cite{CannensonNanoLett2017};
Brennan \emph{et al}., Ref.~\protect\cite{BrennanJACS2017}.}
\end{figure}

The Kane parameter $\Ep$ of CsPbBr$_{3}$ and CsPbI$_{3}$ has not
been measured directly. An estimate of $\Ep$ can be made by assuming
that the contribution to $m_{e}^{*}$ and $m_{h}^{*}$ from remote
bands is zero, which in the $8\times8\ \mathbf{k}\cdot\mathbf{p}$
model implies \cite{[{This follows by putting $\gamma_e=1$ and $\gamma_h=-1$ in }][{}]EfrosRosen_annurev.matsci2000}
\begin{equation}
\frac{1}{\mu^{*}}=\frac{2}{3}\left(\frac{\Ep}{\Egap}+\frac{\Ep}{\Egap+\Esoc}\right)\,,\label{eq:muandEp}
\end{equation}
where $\Egap$ is the gap energy. The Kane parameter here is defined
by 
\begin{equation}
\Ep=2|\BraOperKet{S}{p_{z}}{Z}|^{2}\,,\label{eq:KaneParameter}
\end{equation}
where $\Ket{S}$ is the Bloch state of the $s$-like band and $\Ket{Z}$
is the $z$-component of the Bloch state of the (spin-uncoupled) $p$-like
band \footnote{The Kane parameter $\Ep$ is sometimes defined to be 1/3 of this value
in the context of the $4\times4$ $\kp$ model. See, for example,
Ref.~\protect\cite{YangACSEnergyLett2017}.}. Equation~(\ref{eq:muandEp}) can now be solved for $\Ep$. The
corresponding equation \cite{even-14b-sqd,YangACSEnergyLett2017}
for the $4\times4$ $\kp$ model is obtained by allowing $\Esoc\rightarrow\infty$.
The values of $\Ep$ inferred in this way for the two models are summarized
in Table~\ref{tab:parameters}.

We take the view that $\Ep$ is uncertain. A conservative range would
be $10\,\text{eV}\leq\Ep\leq32\,\text{eV}$ for CsPbBr$_{3}$ and
$8\,\text{eV}\leq\Ep\leq26\,\text{eV}$ for CsPbI$_{3}$. Note that
the uncertainty in $\Ep$ is not critical for the calculation of the
energy, since $\Ep$ determines only the rather small $\kp$ corrections
to $\Ehf^{\text{av}}$ and $E_{\text{corr}}$ (see Secs.~\ref{sec:HFApproximation}
and \ref{sec:MBPTFormalism}). For illustrative purposes, we choose
a central value of $\Ep=20$~eV for CsPbBr$_{3}$ in Tables~\ref{tab:HF}
and \ref{tab:correlation}.

An overall assessment of the parameters and the model can be made
by comparing the theoretical single-exciton energy with the energy
of the emission peak \footnote{Note that in NCs of CsPbBr$_{3}$, there is a Stokes shift of the
emission peak (to lower energies than the first absorption peak),
which varies from a few tens of meV at the larger NC sizes $L\agt8$~nm
to about 80~meV at the smaller sizes $L\approx4$~nm; see Ref.~\protect\cite{BrennanJACS2017}.
One might also compare the theoretical single-exciton energy with
the position of the first absorption peak, but the resulting change
in Fig.~\protect\ref{fig:PLpeaks} would not be very great, and
it is harder to obtain experimental values for the absorption peak
because few authors provide explicit peak-fits to their absorption
curves.}, as shown in Fig.~\ref{fig:PLpeaks}. The data in the figure correspond
to a variety of experimental conditions. Most of the measurements
were made at room temperature, although Yin \emph{et al}.\ \cite{YinPRL2017}
(CsPbI$_{3}$) and Canneson \emph{et al}.\ \cite{CannensonNanoLett2017}
(CsPbBr$_{3}$) were at cryogenic temperatures, as were the measurements
\cite{YangACSEnergyLett2017} used to determine our parameters (Table~\ref{tab:parameters}),
which are therefore more appropriate to low temperatures. This explains
part of the apparent small discrepancy at large sizes $L$, as the
bandgap increases at room temperature by about 60~meV (CsPbBr$_{3}$)
to 80~meV (CsPbI$_{3}$) \cite{YangACSEnergyLett2017}. Also, the
measurement of Liu \emph{et al}.\ \cite{LiuACSNano2017} (CsPbI$_{3}$)
is ligand-dependent, as indicated by the multiple data points.

It is clear from Fig.~\ref{fig:PLpeaks} that the contribution of
correlation to the total emission frequency is not significant, but
the role of correlation is greatly enhanced in measurements of the
trion and biexciton shifts, which are discussed in the next section.

\subsection{\label{sec:ClosedShell}Bi-exciton and trion shifts}

Emission from trions or biexcitons in NCs is usually observed to occur
at a slightly lower frequency than from a single exciton \cite{CastanedaACSNano2016,AneeshACSjpcc2017,NakaharaACSjpcc2018,ZungerCorrPRB2001,raino-16-sqd}.
The trion $\Delta_{X^{-}}$ and biexciton $\Delta_{X\!X}$ redshifts,
relative to the single-exciton emission frequency, can be found by
taking the difference of the initial and final energies, 
\begin{eqnarray}
\Delta_{X\!X} & = & 2E_{X}-E_{X\!X}\,,\label{eq:biexcitonShift}\\
\Delta_{X^{-}} & = & E_{X}+E_{1e}-E_{X^{-}}\,,\label{eq:trionShift}
\end{eqnarray}
where $E_{X}$, $E_{X^{-}}$, $E_{X\!X}$, and $E_{1e}$ are the total
energies of the single exciton, the negative trion, the biexciton,
and a single confined electron, respectively. We assume here that
the excitonic systems relax nonradiatively under experimental conditions
before emitting, so that these total energies will be taken to refer
to the ground state. In Eqs.~(\ref{eq:biexcitonShift}) and (\ref{eq:trionShift}),
we have anticipated that the shifts are redshifts by defining $\Delta_{X^{-}}$
and $\Delta_{X\!X}$ as minus the change in energy relative to a single
exciton. Because we assume VB-CB symmetry of effective-mass parameters
(see Table~\ref{tab:parameters}), the positive trion will have an
identical shift to the negative trion, $\Delta_{X^{+}}=\Delta_{X^{-}}$.

\begin{table}
\caption{\label{tab:biTriShifts}Calculated biexciton $\Delta_{X\!X}$ and
trion $\Delta_{X^{-}}$ redshifts in NCs of CsPbBr$_{3}$ with edge-lengths
$4\,\text{nm}\protect\leq L\protect\leq12\,\text{nm}$, assuming the
material parameters in Table~\protect\ref{tab:parameters} (and
$\Ep=20$~eV). The contributions to Eqs.~(\protect\ref{eq:biexcitonShift})
and (\protect\ref{eq:trionShift}) from Hartree-Fock (HF) and correlation
(Corr) are shown separately; $\Delta_{X\!X}$ and $\Delta_{X^{-}}$
are the sum of the HF and Corr terms. Units: meV.}

\begin{ruledtabular}
\begin{tabular}{ldddd}
 & \multicolumn{1}{c}{4~nm} & \multicolumn{1}{c}{6~nm} & \multicolumn{1}{c}{9~nm} & \multicolumn{1}{c}{12~nm}\\
\hline 
HF  & -3.96  & -1.57  & -0.58  & -0.28 \\
Corr  & 18.16  & 14.68  & 11.58  & 9.62 \\
$\Delta_{X\!X}$  & 14.19  & 13.11  & 11.00  & 9.34 \\
\hline 
HF  & -0.05  & 1.03  & 1.41  & 1.47 \\
Corr  & 10.84  & 9.15  & 7.61  & 6.50 \\
$\Delta_{X^{-}}$  & 10.79  & 10.18  & 9.02  & 7.96 \\
\end{tabular}
\end{ruledtabular}

\end{table}

Remarkably, the biexciton {[}Eq.~(\ref{eq:biexcitonShift}){]} and
trion {[}Eq.~(\ref{eq:trionShift}){]} shifts are dominated by intercarrier
correlation effects, as the mean-field contribution largely cancels
\cite{ZungerCorrPRB2001}. This phenomenon for a NC of CsPbBr$_{3}$
is illustrated in Table~\ref{tab:biTriShifts}. Note that for large
edge-lengths $L\agt7$~nm, the cancelation of the HF contribution
is more complete for the biexciton than for the trion, with the reverse
being true for the smaller edge-lengths tabulated. The final shifts
$\Delta_{X\!X}$ and $\Delta_{X^{-}}$ have a quite weak size dependence.
This can be understood by noting that a Coulomb matrix element scales
approximately as $\BraOperKet{ab}{g_{12}}{rs}\sim1/L$, while the
energy denominator in Eq.~(\ref{eq:D2general}) scales approximately
as $\omega_{a}+\omega_{b}-\omega_{r}-\omega_{s}\sim1/L^{2}$ owing
to the confinement effect~(\ref{eq:eigsCube}), so that $E^{(2)}$
is approximately independent of $L$. In fact, both $\Delta_{X\!X}$
and $\Delta_{X^{-}}$ become slightly larger at the smaller sizes
in Table~\ref{tab:biTriShifts}, an effect that has been observed
experimentally in perovskite NCs \cite{CastanedaACSNano2016}.

\begin{figure}[tb]
\includegraphics[scale=0.85]{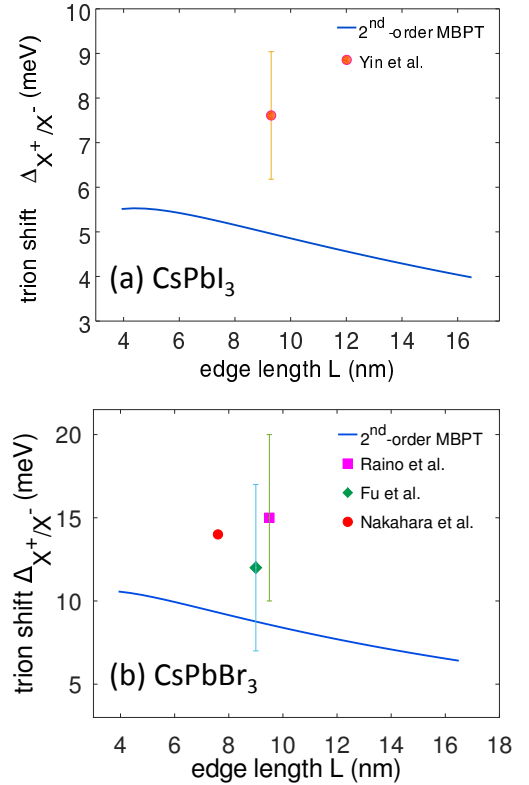}

\caption{\label{fig:trionShift}Measured trion redshift $\Delta_{X^{-}}$ of
NCs of (a) CsPbI$_{3}$ and (b) CsPbBr$_{3}$ (squares/circles/diamonds).
Solid line: theory (second-order MBPT). Yin \emph{et al}., Ref.~\protect\cite{YinPRL2017};
Rain{\`o} \emph{et al}., Ref.~\protect\cite{raino-16-sqd}; Fu
\emph{et al}., Ref.~\protect\cite{FuNanoLett2017}; Nakahara \emph{et
al}., Ref.~\protect\cite{NakaharaACSjpcc2018}.}
\end{figure}

\begin{figure}
\includegraphics[scale=0.85]{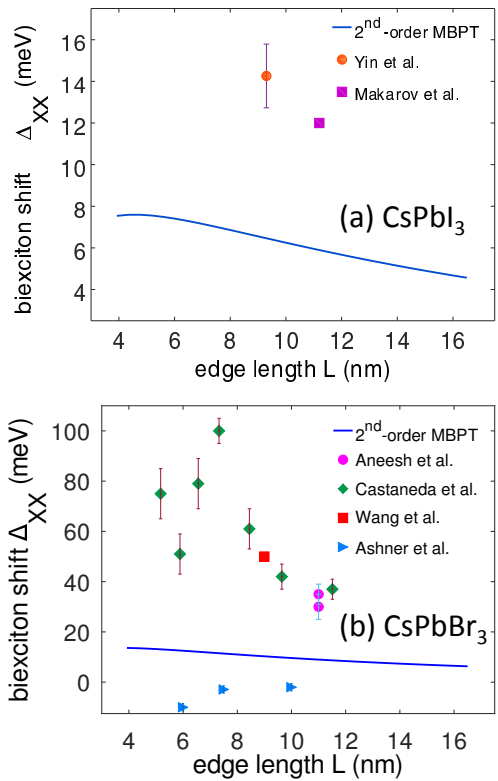}

\caption{\label{fig:biexcitonShift}Measured biexciton redshift $\Delta_{X\!X}$
of NCs of (a) CsPbI$_{3}$ and (b) CsPbBr$_{3}$ (triangles/squares/circles/diamonds).
Solid line: theory (second-order MBPT). Yin \emph{et al}., Ref.~\protect\cite{YinPRL2017};
Makarov \emph{et al}., Ref.~\protect\cite{makarov-16-sqd}; Aneesh
\emph{et al}., Ref.~\protect\cite{AneeshACSjpcc2017}; Castaneda
\emph{et al}., Ref.~\protect\cite{CastanedaACSNano2016}; Wang \emph{et
al}. Ref.~\protect\cite{WangAdvMater2015}; Ashner \emph{et al}.,
Ref.~\protect\cite{ashner-19-sqd}.}
\end{figure}

Our theoretical trion shifts are compared with the available experimental
data in Fig.~\ref{fig:trionShift}. The agreement with the trion
data is fair, although the theoretical values are perhaps systematically
too small (by a factor of order 1.3--1.8). Turning to the data on
the biexciton shift, shown in Fig.~\ref{fig:biexcitonShift}, we
see that a similar comment holds for CsPbI$_{3}$, where the theoretical
values are smaller than the few available measurements by a factor
of about 1.8--2.0. The situation is rather unclear for the biexciton
shift in CsPbBr$_{3}$, however, where there are more data available.
The measured values of $\Delta_{X\!X}$ for CsPbBr$_{3}$ range from
large redshifts \cite{CastanedaACSNano2016} of about 40--100~meV
(which is comparable to the HF Coulomb energy given in Table~\ref{tab:HF})
to a recently reported small blueshift \cite{ashner-19-sqd}, of order
$\Delta_{X\!X}=-2$~meV for $L\approx10$~nm. Our second-order MBPT
approach predicts a redshift for all sizes considered for both CsPbBr$_{3}$
and CsPbI$_{3}$, with a value $\Delta_{X\!X}=10$~meV for CsPbBr$_{3}$
for $L\approx10$~nm.

Before commenting on the experimental data, let us first review some
leading sources of theoretical error in our second-order MBPT approach.
These are:

(i) \emph{Correction terms due to fine-structure splittings}. We have
neglected the fine structure (FS) of the excitonic states, basing
our formalism on a configuration-averaged approach~(\ref{eq:D2av}),
which yields the center of gravity of the FS multiplet. FS splittings
in emission lines of inorganic perovskite NCs are observed to vary
from several hundred $\mu$eV (e.g., Ref.~\cite{YinPRL2017}) to
a few meV (e.g., Ref.~\cite{BeckerNatLett2018}). Single-dot spectroscopy
reveals that they can vary quite markedly from dot to dot, both in
magnitude and sometimes also in the number of FS components observed
\cite{FuNanoLett2017,YinPRL2017,BeckerNatLett2018}. In the few cases
that FS splittings have been observed experimentally in the measurements
relevant to Figs.~\ref{fig:trionShift} and \ref{fig:biexcitonShift},
the shifts plotted in the figures correspond to the values obtained
by averaging over the FS (e.g., Ref.~\cite{YinPRL2017}). Because
of this and the relatively small size of the FS splittings, the error
in Figs.~\ref{fig:trionShift} and \ref{fig:biexcitonShift} due
to FS seems likely to be at the level of 1--2~meV or less.

Let us consider the role of the FS of the single exciton in greater
detail. In perovskite NCs, the ground-state $1S_{e}$-$1S_{h}$ single
exciton consists of electron and hole states with angular momentum
$F=1/2$, in the notation of Eq.~(\ref{eq:4compState}), which can
couple to a total angular momentum $F_{\text{tot}}=0$ or 1 (singlet
or triplet, respectively). The triplet state has an allowed electric-dipole
radiative decay and is a bright exciton state; the singlet is a dark
state \cite{FuNanoLett2017,BeckerNatLett2018}. Similarly, the ground-state
biexciton in perovskite NCs has closed-shell electron and hole states,
$1S_{e}^{2}$-$1S_{h}^{2}$, which must therefore couple to $F_{\text{tot}}=0$,
and the negative trion has a $1S_{e}^{2}$-$1S_{h}$ ground state
with $F_{\text{tot}}=1/2$. From selection rules, the allowed biexciton
emission must proceed via the bright single-exciton state, $X\!X_{0}\rightarrow X_{1}$,
where the subscript indicates the value of $F_{\text{tot}}$.

Now, the center of gravity of the bright-dark FS multiplet in the
single exciton is given by 
\begin{equation}
\bar{E}_{X}=(1/4)E_{X_{0}}+(3/4)E_{X_{1}}\,,\label{eq:CGravityEX}
\end{equation}
from which it follows that $E_{X_{1}}-\bar{E}_{X}=\Delta_{10}/4$,
where $\Delta_{10}=E_{X_{1}}-E_{X_{0}}$ is the bright-dark FS splitting.
(We are assuming that any FS in the bright state, which is due to
nonspherical or noncubic symmetry-breaking interactions \cite{FuNanoLett2017,BeckerNatLett2018,ben-aich-19-sqd,SercelNanoLett2019},
has been experimentally averaged.) For the biexciton, the configuration-averaged
energy $\bar{E}_{X\!X}=E_{X\!X_{0}}$, since there is only one state.
Therefore, the observed biexciton shift is given by 
\begin{equation}
\Delta_{X\!X}=2E_{X_{1}}-E_{X\!X_{0}}=(2\bar{E}_{X}-\bar{E}_{X\!X})+\Delta_{10}/2\,,\label{eq:DxxObserved}
\end{equation}
and we see that our calculated result in Fig.~\ref{fig:biexcitonShift}
acquires a correction term $\Delta_{10}/2$. An analogous argument
leads to a correction term $\Delta_{10}/4$ for the trion shift in
Fig.~\ref{fig:trionShift}. Since $|\Delta_{10}|$ is expected to
be of order a few meV \cite{SercelNanoLett2019}, we conclude again
that any error in Figs.~\ref{fig:trionShift} and \ref{fig:biexcitonShift}
from this source is likely to be of order at most 1--2~meV.

(ii) \emph{Uncertainty in the value of the dielectric constant}. The
biexciton $\Delta_{X\!X}$ and trion $\Delta_{X^{-}}$ shifts are
dominated by correlation or $E^{(2)}$, so that they are both approximately
proportional to $1/\epsin^{2}$, where $\epsin$ is the dielectric
constant of the material (\ref{eq:lrcoul}). However, a more complete
treatment of dielectric effects than considered in the present paper
would take into account the space- and frequency-dependent bulk dielectric
function $\varepsilon(\mathbf{k},\omega)$. In the instantaneous approximation
$\omega=0$, the dielectric constant $\epsin$ in Eq.~(\ref{eq:lrcoul})
would then be replaced by a space-dependent function $\varepsilon(\mathbf{r}_{1},\mathbf{r}_{2})$.
A more general treatment including also the frequency-dependence of
$\varepsilon(\mathbf{k},\omega)$ would require a retarded Coulomb
interaction (and, for example, the use of Feynman propagators \cite{Mahan}).
Inorganic perovskites present the complication that the dielectric
function is rapidly varying; for instance, the effective and optical
dielectric constants given in Table~\ref{tab:parameters} are quite
different.

Another dielectric effect, which we have neglected here, arises from
the mismatch of the dielectric constant of the NC with that of the
surroundings, which modifies the effective LR Coulomb interaction
to take account of polarization charges induced at the dielectric
boundary \cite{karpulevich-19}.

In our calculations, we have assumed a dielectric constant $\epsin=\epseff$,
where $\epseff$ is derived from the measured binding energy of the
bulk exciton (see Sec.~\ref{sec:parameters}). Formally, $\epseff$
corresponds to length scales of order the Bohr radius $k\sim\pi/a_{B}$
and to a frequency $\omega\approx0$, since the exciton binding energy
is dominated by the direct Coulomb energy (\ref{eq:HFdir}) and (\ref{eq:Ecoul}),
in which the energy flowing through the Coulomb propagator in the
Feynman rules is zero. The second-order energy $E^{(2)}$, on the
other hand, involves a nonzero average excitation energy $\delta\omega_{\text{av}}=\langle\omega_{a}+\omega_{b}-\omega_{r}-\omega_{s}\rangle$
in Eq.~(\ref{eq:D2av}), which implies a nonzero average energy flowing
through the Coulomb propagators. We find $\delta\omega_{\text{av}}\approx0.1$--0.6~eV
for $4\text{ nm}\leq L\leq12\text{ nm}$ for NCs of CsPbBr$_{3}$.
In addition, although the size of our NCs is comparable to the Bohr
radius (see Sec.~\ref{sec:parameters}), this is not exactly true.
It follows that the appropriate value of the dielectric constant $\epsin$
to use in calculations of $E^{(2)}$ might differ from $\epseff$.
For instance, it seems likely that the frequency-dependence will shift
the appropriate value of $\epsin$ from $\epseff$ toward a slightly
smaller value, closer to $\epsopt$ (see Table~\ref{tab:parameters}).
This would increase $E^{(2)}$ and could explain part of the discrepancy
between theory and experiment observed in Figs.~\ref{fig:trionShift}
and \ref{fig:biexcitonShift}(a).

(iii) \emph{Higher-order correlation}. Usually in atoms and molecules,
$E^{(2)}$ underestimates the all-order correlation energy \cite{LindgrenMorrison,ShavittBartlett}.
Unfortunately, it is hard to estimate the higher-order correlation
$E^{(3+)}=E^{(3)}+E^{(4)}+\ldots$ without explicit calculation, although
we note that typical values of $E^{(3+)}$ for atoms and molecules
can vary up to 25\% or so of $E^{(2)}$, depending on the system.
Each order of MBPT brings in one extra Coulomb interaction $g$ and
an energy denominator $\Delta\epsilon$, which scale approximately
as $g/\Delta\epsilon\sim L$. Therefore the contribution of higher-order
MBPT is expected to become more important for larger dots, and this
could explain a large part of the discrepancies noted in Figs.~\ref{fig:trionShift}
and \ref{fig:biexcitonShift}(a) for the case of intermediate confinement
encountered in perovskite NCs.

Table~\ref{tab:correlation} makes it clear that the calculations
of $E^{(2)}$ for the trion and the biexciton are very closely related.
The term $E_{ee}^{(2)}$ in Eq.~(\ref{eq:E2decomposed}) can be seen
to have almost the same value for each. This happens because both
systems contain two electrons, so that the configuration-averaging
factors in Eq.~(\ref{eq:D2av}) are the same (although the basis
sets differ slightly, because different states are occupied in the
HF potential of the two systems). Similarly, most of the difference
in the other two terms $E_{eh}^{(2)}$ and $E_{hh}^{(2)}$ in Table~\ref{tab:correlation}
is due simply to the different configuration-averaging weights for
the trion and biexciton. Because of this, we expect that the errors
in $\Delta_{X\!X}$ and $\Delta_{X^{-}}$ due to both dielectric effects
{[}(ii) above{]} and omitted higher-order MBPT {[}(iii) above{]} should
be comparable. The trion data in Fig.~\ref{fig:trionShift} can therefore
serve as an additional check on the biexciton data in Fig.~\ref{fig:biexcitonShift}.

Turning to the experimental data, we note first that it is useful
to distinguish between measurements on single dots at cryogenic temperatures
(e.g., using time-resolved photoluminescence) and high-temperature
measurements on ensembles of NCs (e.g., using transient absorption).
The low-temperature measurements typically give narrow well-separated
peaks, from which the shifts can be extracted directly, while the
high-temperature measurements typically require extensive fits to
side-features on overlapping peaks, or other indirect analysis methods.
Low-temperature single-dot measurements have been performed on the
trion (Fig.~\ref{fig:trionShift}) by Fu \emph{et al}.\ \cite{FuNanoLett2017}
for CsPbBr$_{3}$ and by Yin \emph{et al}.\ \cite{YinPRL2017} for
CsPbI$_{3}$, and the latter also measured the biexciton shift for
CsPbI$_{3}$ (Fig.~\ref{fig:biexcitonShift}a). No low-temperature
measurements are available of the biexciton shift in CsPbBr$_{3}$.

We observe that our agreement with all these low-temperature measurements
in the trion shift (Fig.~\ref{fig:trionShift}) is fair. Based on
this, and the observation that the theoretical errors for the trion
and the biexciton shift are expected to be similar, we believe that
the present results provide quite strong theoretical evidence that
the ground-state biexciton shift in NCs of CsPbBr$_{3}$ is a redshift
of order $\Delta_{X\!X}=10$--20~meV for $L\approx12$~nm (after
allowing for a phenomenological increase in the second-order MBPT
values given in Table~\ref{tab:biTriShifts} by a factor of up to
2).

According to Shulenberger \emph{et al}.\ \cite{shulenberger-19-sqd},
who performed experiments on NCs of CsPbBr$_{3}$, the fast red-shifted
features often attributed to biexciton emission are actually an artifact
of the exposure of the sample to air, which they claim causes the
formation of larger bulk-like particles in the ensemble with a red-shifted
single-exciton peak. Shulenberger \emph{et al}.\ \cite{shulenberger-19-sqd}
placed an upper limit on the true biexciton shift of 20~meV, which
is consistent with our theoretical prediction. However, the same group
later inferred \cite{ashner-19-sqd} a small biexciton \emph{blueshift}
of order $\Delta_{X\!X}=-2$~meV for $L\approx10$~nm after extensive
data fitting, which seems to be inconsistent with our theoretical
value.

Another experimental issue is whether the biexciton has truly relaxed
to the ground state, as we have assumed in our calculation. Yumoto
\emph{et al}.\ \cite{yumoto-18-sqd} studied `hot' biexcitons in
a transient absorption experiment on NCs of CsPbI$_{3}$ by observing
the induced absorption signal immediately after the pump excitation.
They concluded that a hot biexciton, composed of one exciton at the
band edge and a second excited exciton, had a substantially increased
exciton-exciton interaction. They found that $\Delta_{X\!X}$ for
CsPbI$_{3}$ could be as large as 60~meV for excitation energies
$E_{\text{ex}}$ of the second exciton of order $E_{\text{ex}}\agt0.3$~eV.

Finally, we note that Makarov \emph{et al}.\ \cite{makarov-16-sqd}
measured $\Delta_{X\!X}=12$~meV for NCs of CsPbI$_{3}$ and obtained
almost the same value $\Delta_{X\!X}=11$~meV for NCs of the mixed
perovskite CsPbI$_{1.5}$Br$_{1.5}$, which would imply a biexciton
shift for CsPbBr$_{3}$ in agreement with our theoretical value.

\section{\label{sec:Conclusions} Conclusions}

We have presented a calculation of the trion and biexciton shifts
in NCs of CsPbI$_{3}$ and CsPbBr$_{3}$ using second-order MBPT.
The agreement with the available data for the biexciton shift in CsPbI$_{3}$
and the trion shift in both CsPbI$_{3}$ and CsPbBr$_{3}$ is fair,
although the theoretical values seem to be systematically slightly
smaller than the measurements, a result that can be plausibly understood
in terms of a slightly overestimated dielectric constant and omitted
higher-order terms in MBPT. After taking this level of agreement between
theory and experiment into account, we infer that the ground-state
biexciton shift in NCs of CsPbBr$_{3}$ is a redshift with a value
of order 10--20~meV (for a size $L=12$~nm). This value is intermediate
in the large range of measured values for CsPbBr$_{3}$.

The theoretical approach used can be improved in various ways in future
work. It is possible to include higher-order MBPT for excitonic systems
with few carriers by means of all-order procedures such as full configuration
interaction \cite{ShavittBartlett}. A better understanding of the
dielectric function in perovskites could perhaps be obtained using
\emph{ab intio} atomistic codes \cite{katan-19-sqd}. The LR Coulomb
interaction~(\ref{eq:lrcoul}) can also be generalized to take account
of the dielectric mismatch with the surrounding medium \cite{karpulevich-19}.
Although envelope-function methods naturally work better on larger
NCs, where atomistic effects are relatively less significant, an important
atomistic effect can be included straightforwardly by assuming a diffuse
finite surface barrier instead of an abrupt infinite barrier~(\ref{eq:sphericalWell}).
Also, explicit nonspherical corrections for the cubic NC shape could
be added as perturbations.

Finally, it should be possible to generalize the methods presented
here to study hot biexcitons, in which one or both excitons are excited.
It would also be interesting to study thermal effects on the biexciton
shift at high temperature, a regime that is more relevant to the conditions
found in practical devices. The present paper assumes that the excitonic
systems are in their quantum ground state, so that it is perhaps natural
to expect better agreement with the low-temperature data.

\begin{acknowledgments}
The authors would like to thank Sum Tze Chien for helpful discussions.
TN and SB gratefully acknowledge Fr{\'e}d{\'e}ric Schuster of the
CEA's PTMA for support. 
\end{acknowledgments}

% BIBLIOGRAPHY

\bibliography{Correlation_purged}

\end{document}